\documentclass[11pt]{article}
\usepackage[utf8]{inputenc}
\usepackage[T1]{fontenc}
\usepackage{lmodern}
\usepackage[margin=1in]{geometry}
\usepackage{amsmath,amssymb}
\usepackage{graphicx}
\usepackage{tikz}
\usetikzlibrary{positioning,arrows.meta,calc}
\usepackage{booktabs}
\usepackage{hyperref}
\usepackage{xcolor}
\usepackage[numbers]{natbib}
\usepackage{enumitem}
\usepackage{placeins}
\usepackage{url}
\usepackage{microtype}
\hypersetup{colorlinks=true, linkcolor=blue, citecolor=blue, urlcolor=blue}

\title{Every Alarm Has a Cause: An Open Generator of Telecom Alarm Storms
with Exact Root Causes, Bound to NORIA-O, RFC 8632 and TMF642}
\author{Madhulatha Mandarapu\thanks{madhulatha@samyama.ai} \and
        Sandeep Kunkunuru\thanks{sandeep@samyama.ai}}
\date{VaidhyaMegha Private Limited, India\\[2pt]\url{https://samyama.ai/}\\[8pt]
      October 2026}

\begin{document}
\maketitle

\begin{abstract}
Alarm root-cause analysis is studied on data that cannot be shared. Operator logs are
confidential, so what the field publishes is a causal graph between alarm \emph{types}
rather than per-incident causes, a fixed scenario set, gated, or never released. Two recent
benchmarks carry per-incident root causes at scale; neither states a download location or a
licence.

We give a generator rather than a dataset. It runs on real backbone topologies from the
Internet Topology Zoo and records the exact root cause \emph{and the propagation path} of
every alarm --- a by-product of generation, not a labelling pass. Every emitted
field binds to a term in NORIA-O, RFC~8632, RFC~8345 or TMF642, and an unknown term fails
the run. We ship fixed configurations, splits, a scorer and nine baselines.

A one-line rule --- rank the device that raised the earliest critical alarm --- takes top-1
at $0.650$ against $0.575$ for our own temporal method, and wins $13$ of $15$ runs over three
backbones and five seeds. An LLM given the same view reaches $0.600$, the same model
allowed to query the graph $0.625$, and the rule wins on real operator alarms
too, at $0.773$: a property of alarm streams, not of our generator. One fault class
separates the table: when a site loses power its PoP reports nothing, so every method that
ranks alarmed devices scores $0.000$ or near it, while the two that walk the topology to a
silent candidate take $1.000$ and $0.833$.
\end{abstract}

\section{Introduction}
\label{sec:intro}

A network operations centre sees a storm of alarms and must name the one resource that
caused them. Research on this problem is evaluated on data nobody else can obtain, because
an operator's alarm log carries its customers, its topology and its failures. The
consequence is not that the work is wrong but that it is incomparable: two methods
reporting $0.7$ on two private logs have not been compared.

The resources the field has published do not close that gap, and
Section~\ref{sec:related} goes through them one at a time. Briefly: the most-used open
alarm data gives a causal graph between alarm \emph{types}, with no incident identifier and
no topology~\citep{qiao2023structural,gcastle_dataset}; the benchmarks that do carry a
cause per incident are either announced without a download~\citep{tnautorca2025,telerca2026}
or gated with terms that contradict their own licence tag~\citep{telelogs}; and the open
RCA benchmarks that are genuinely downloadable are
microservice systems, not telecom alarms~\citep{rcaeval2024,lemmarca2024}. One paper
advertises its data at a repository that contains a single 378-byte
README~\citep{zhang2021influence}, and is still cited as a data source five years later.

\paragraph{A generator, not a dataset.} We take the other route. A generator sidesteps
confidentiality in a way no release process can, and it gives something a released log
cannot: the ground truth is not an annotation added afterwards by an expert who was not
there, it is what the generator did. That distinction matters most for the
\emph{propagation path}. An annotator can reasonably name the resource that caused an
incident; reconstructing which alarm caused which other alarm, through what, is a different
task, and the largest prior benchmark states plainly that it does not have
one~\citep{telerca2026}.

\paragraph{The claim.} We do not claim the first per-incident root causes for telecom
alarms. Two 2025--2026 benchmarks claim that, and at a realism we cannot match --- real
production networks, expert annotation, hundreds of thousands of
alarms~\citep{telerca2026,tnautorca2025}. Our claim is narrower:

\begin{quote}
No \emph{downloadable, openly licensed} resource gives per-incident root causes
\emph{and propagation paths} for telecom alarms at scale, and no open generator exists.
\end{quote}

Both halves were checked against the sources rather than assumed, on 2026-10-21, and the
check is published with the paper.

\paragraph{Contributions.}
\begin{enumerate}[leftmargin=1.4em,itemsep=2pt]
\item A generator of telecom alarm storms that records the root cause and the full
      propagation path of every alarm, parameterised by topology and seeded
      (Section~\ref{sec:generator}).
\item A binding of every emitted type and field to NORIA-O, RFC~8632, RFC~8345 or TMF642,
      enforced at generation time (Section~\ref{sec:binding}). NORIA-O itself aligns to
      ITU-T X.733/X.721 and TMF638 and references neither RFC~8632 nor TMF642; the binding
      to those two is ours, not inherited.
\item A benchmark --- fixed configurations, splits, a scoring script, and baselines we did
      not design (Sections~\ref{sec:benchmark}--\ref{sec:baselines}).
\item Nine scored baselines --- four one-line rules, a fitted topological Hawkes process,
personalised PageRank, a language model given the same view, the same model allowed to
query the graph, and a control that isolates what the winning rule actually uses: severity,
not time (Section~\ref{sec:results}).
\item A realism study against public real alarm data, including the test that decides
whether the resource is useful: the ranking of methods on generated data is the ranking on
real data (Kendall $\tau = 1.0$ over four rules on $4{,}539$ real bursts), while three
realism metrics show the generated stream matches real alarms on type vocabulary and
chaining and diverges on time, arriving in burstier volleys.
\end{enumerate}

\section{Related work}
\label{sec:related}

Every row of Table~\ref{tab:related} was read from the source on 2026-10-21 rather than
from a secondary description, and the record of what was checked ships with the paper.
Two things we expected to be true were not, and both are stated here rather than quietly
corrected.

\begin{table}[t]
\centering\small
\begin{tabular}{@{}llll@{}}
\toprule
Resource & Obtainable & Labels & Topology \\
\midrule
PCIC 2021~\citep{gcastle_dataset} & yes, MIT & alarm-\emph{type} graph & none \\
Influence-based~\citep{zhang2021influence} & no & per-transaction (unreleased) & none \\
TN-RCA530~\citep{tnautorca2025} & announced only & per-incident & per-scenario KG \\
TeleRCA~\citep{telerca2026} & announced only & per-incident, no path & none published \\
TeleLogs~\citep{telelogs} & gated & 8 fixed categories & none \\
RCAEval~\citep{rcaeval2024} & yes & per-case service & microservice \\
LEMMA-RCA~\citep{lemmarca2024} & yes, ND & per-case pods & microservice \\
OpenRCA~\citep{openrca2025} & yes & per-case & cloud telemetry \\
\midrule
\textbf{This work} & \textbf{yes, open} & \textbf{per-incident $+$ path} & \textbf{RFC 8345} \\
\bottomrule
\end{tabular}
\caption{Public resources for alarm root-cause analysis. ``Announced only'' means the
paper states the resource is released and contains no repository, download or licence.}
\label{tab:related}
\end{table}

\paragraph{The closest prior work is TeleRCA, and it is better than ours on realism.}
TeleRCA~\citep{telerca2026} gives 10,922 production incidents, 485,681 alarm events and
5,621 network resources, each incident carrying an expert annotation naming the root-cause
resource. That is a real network and we do not have one. We differ on three things its own
paper establishes: it states twice that it has \emph{no ground-truth propagation graph},
and reports no graph-reconstruction metric in consequence; it publishes no topology or
inventory; and although it says four times that it is released, the paper contains no
repository, no download and no licence. TN-RCA530~\citep{tnautorca2025} is closer to us on
topology --- each of its 530 scenarios is a knowledge graph --- and is likewise announced
without a download or a dataset licence.

\paragraph{The most-used open alarm data labels types, not incidents.} The Huawei PCIC 2021
data~\citep{qiao2023structural} is live under MIT, though not at the URL its paper
prints~\citep{gcastle_dataset}: 34,838 alarms, 18 alarm types, 55 devices over 243 days.
Its ground truth is an $18\times18$ adjacency between alarm \emph{types}. There is no
incident identifier, no storm grouping, no root-cause column and no topology. It is a good
realism reference and not a source of per-incident labels. The nearest open artefact to
such labels is a 49-row prior shipped with a later competition sample, whose ``root cause''
is itself an alarm type.

\paragraph{Availability is the field's recurring failure, and it is the argument for a
generator.} \citet{zhang2021influence} states that its data and code are available at a
named repository; that repository holds one 378-byte README saying the release is going
through internal procedures. The paper is still cited as a data source. TeleRCA and
TN-RCA530 repeat the pattern. A generator is not a better dataset --- it is a way of not
depending on someone else's release process.

\paragraph{Open RCA benchmarks exist, for other systems.} RCAEval~\citep{rcaeval2024} and
LEMMA-RCA~\citep{lemmarca2024} are downloadable and well constructed, and are microservice
and water-treatment systems rather than telecom alarms; LEMMA-RCA carries CC~BY-ND on its
own site, which would block derived works. OpenRCA~\citep{openrca2025} has a domain named
\emph{Telecom}, which invites confusion: it is cloud observability telemetry --- KPIs,
traces and logs of enterprise systems --- not NOC alarms, and the repository states no
licence for the data.

\paragraph{No generator on the Internet Topology Zoo.} We found none that emits alarms with
labelled root causes. The closest prior work describes an optical alarm-propagation
simulator with an ITU-T G.798 rule engine producing alarm correlation trees; no code
repository was found.

\section{The generator}
\label{sec:generator}
Topologies are real backbones taken from the Internet Topology
Zoo~\citep{topologyzoo}, which is CC~BY~4.0.
A backbone's GraphML gives the PoPs and the links between them, and those are the only
real objects in the output; everything below carries \texttt{synthetic: true}.

\paragraph{Inventory and services.} Each PoP gets two core routers, two or three
aggregation routers dual-homed to both, and six to twelve access or radio sites homed to
aggregation, then cards, then ports, then cabling. Backbone links land on core-router
ports, so the real topology constrains the synthetic one rather than sitting beside it.
Services follow the inventory: mobile data per radio site, broadband per access site, and
enterprise VPNs across PoPs with a primary and a backup path. Customers carry a market
segment and nothing else --- there is no personal data to anonymise.

\paragraph{Faults and propagation.} There are $7$ fault types: card, device, port and
software failure, fibre cut, site power loss, and a configuration change, which is always
preceded by a \texttt{Change} whose outcome is \emph{failed}. What follows from a fault is
computed from the topology, not scripted: a dead card takes its ports down, the far end of
each down port loses signal, a breadth-first search from the NOC's own PoP over surviving
links decides which devices still have a path home, and a service whose endpoint is
unreachable is out. Redundancy is what stops a failure spreading, so dual homing and two
core routers per PoP are what make most faults local and a few of them storms.

\paragraph{Alarms.} Each of those states raises an alarm on the resource that would report
it, with a per-mechanism severity and a per-mechanism emission probability: the root alarm
when the root can still speak, loss of signal at the far ends, an unreachable-device alarm
after the next poll, a service alarm per lost service, neighbour-cell link failures on live
radio sites near a dead one, and congestion where traffic reroutes. The dataset uses $4$
X.733 event types and $21$ NORIA-O probable causes over $4$ RFC~8632 severities.

\paragraph{What makes it hard is in the config, not in a script.} The root resource often
raises no alarm at all: in $50$ of $70$ faults it does, and in the rest the method has to
infer the root from what its neighbours said. A site power loss never raises one --- the
PoP it kills cannot report itself, which is also why the fault model's fifth event type,
\texttt{environmentalAlarm}, appears nowhere in the data. On top of that, alarms go
missing, flap through raise/clear cycles, duplicate, and carry per-device clock skew drawn
once per device, so the order alarms arrive in is not the order the faults happened in.
Background alarms arrive throughout, unrelated to any fault. Two faults can overlap inside
one PoP. Every one of those is a knob with a number in the config, which is what lets the
ablation in Section~\ref{sec:results} say which of them the methods are actually reading.

Section~\ref{sec:ablation} removes each of these in turn and measures what it was worth.
Alarms that never arrive are worth the most by a wide margin; mis-grouping is worth the
least, and was worth nothing at all before the arrival times were fixed.

\paragraph{The distribution, not the mean.} Alarms per fault have a median of $3$ and a
90th percentile of $41.1$, with a maximum of $103$. Most faults are quiet and a few are
storms, so a method evaluated at the mean is evaluated at neither end.

\paragraph{What is recorded.} For every injected fault the generator writes the fault, its
onset and repair, the resources it hit, the alarms it caused and the window it spans. This
is the ground truth, and it is a by-product of generation rather than a labelling pass.

\begin{figure}[htbp]
\centering
\begin{tikzpicture}[
  every node/.style={font=\small},
  gstep/.style={draw, rounded corners=2pt, align=center, inner sep=5pt,
               text width=4.6cm, minimum height=9mm},
  greal/.style={gstep, fill=black!8},
  gout/.style={draw, thick, rounded corners=2pt, align=left, inner sep=5pt,
              text width=5.3cm, fill=black!4},
  ar/.style={-{Latex[length=2.2mm]}, thick},
  glab/.style={font=\scriptsize\itshape, align=left}
]
\node[greal] (zoo) at (0,0) {real backbone\\(Topology Zoo GraphML)};
\node[gstep] (inv) at (0,-1.45) {inventory, services, customers};
\node[gstep] (split) at (0,-2.9) {history / eval split, before generation};
\node[gstep] (fault) at (0,-4.35) {fault injection (7 types)};
\node[gstep] (prop) at (0,-5.8) {propagation computed over the topology};
\node[gstep] (emit) at (0,-7.25) {alarm emission, with the noise knobs};

\foreach \a/\b in {zoo/inv, inv/split, split/fault, fault/prop, prop/emit}
  \draw[ar] (\a) -- (\b);

\node[gout] (gt) at (7.1,-5.55) {\textbf{ground truth}\\[2pt]
  fault, onset, repair, resources hit, propagation path, window};
\node[gout] (stream) at (7.1,-7.5) {\textbf{alarm stream}\\[2pt]
  all a method sees on eval, plus the static topology};

\draw[ar] (prop) -- (gt);
\draw[ar] (emit) -- (stream);

\node[glab] at (7.1,-3.4) {\parbox{5.3cm}{The ground truth forks off the
  propagation step. It is a by-product of generation, not a labelling pass
  over the stream below it.}};
\end{tikzpicture}
\caption{How a run is produced. Only the top box is real; everything below it
carries \texttt{synthetic:\ true}. The two outputs are written by the same pass,
which is why every alarm's cause and path are exact rather than inferred.}
\label{fig:pipeline}
\end{figure}
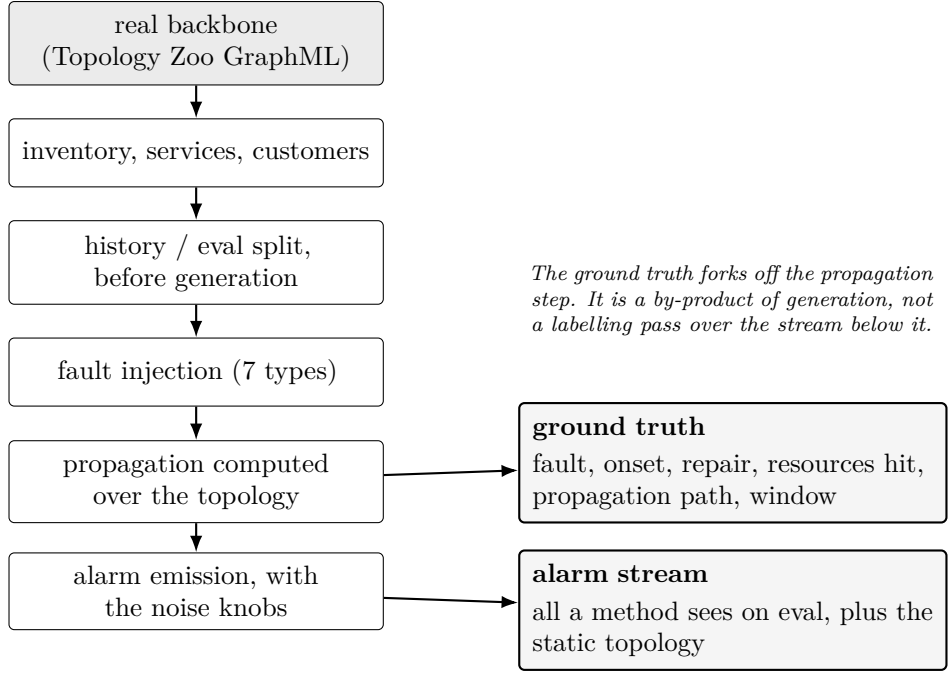

\paragraph{Splits.} Scenarios are split into \emph{history} and \emph{eval} before
generation, not after, and each scenario owns its own window, so the two never interleave
in time. A method may use everything about history --- its alarms, its incidents, its
resolved tickets, its changes, and the ground truth --- to fit whatever it likes. On eval
it sees the static topology and the incident's alarms, and nothing else. The GÉANT small
configuration has $30$ history faults and $40$ eval faults in three scenario shapes:
a single fault, a storm, and two faults overlapping inside one PoP.

The split is only a split if a method cannot see across it, which is a property of the
method and not of the data. The one baseline here that fits anything --- the topological
Hawkes process in Section~\ref{sec:baselines} --- is given a hard cutoff at the onset of
the first eval incident, and a test asserts that every alarm it was trained on falls below
that cutoff and that the cutoff excludes a non-zero number of them. A cutoff that excludes
nothing proves nothing.

\section{Standards binding}
\label{sec:binding}
Every emitted type and field resolves to a term in one of four published vocabularies:
NORIA-O~\citep{noria2024}, the alarm-management YANG model of RFC~8632~\citep{rfc8632},
the network-topology model of RFC~8345~\citep{rfc8345}, and the TM Forum alarm
API TMF642~\citep{tmf642}. Generation fails on an unknown term rather than emitting it.
The binding is a build step rather than a convention. Each vocabulary is converted to a
graph of its own terms --- an OWL/SKOS pass for NORIA-O, a YANG pass for the two RFCs, an
OpenAPI pass for TMF642 --- and a catalog entry records where each build came from. The
generator then declares what it needs: for each node label, the ontology classes it claims
to be, and for each enumerated field, the vocabulary and the owning term the values must
come from. At generation time every label and every enumeration value is resolved against
those builds, $11$ labels and $4$ vocabularies in this configuration, and an unresolved
term raises rather than being emitted. The release carries \texttt{binding.json}: the terms
used, and the build files they came from with their sha256, so a reader can check the
vocabulary a value claims without trusting the paper.

The grounding also travels in the graph itself, as one \texttt{SchemaLabel} node per label
with a \texttt{GROUNDED\_IN} edge to each ontology concept, rather than as a table in a
README. A query can therefore ask what an \texttt{Alarm} is in NORIA-O's terms.

\paragraph{What is ours and what is inherited.} NORIA-O's own alarm vocabulary aligns to
ITU-T X.733/X.721 and to TMF638 Service Inventory; neither it nor its IETF
draft~\citep{noria_draft} references RFC~8632 or TMF642. Binding to those two is this
paper's contribution rather than an alignment we inherited, and we state it so that a
reader who opens the ontology does not have to discover it.

\paragraph{Licence.} NORIA-O is BSD-4-Clause, which carries an advertising clause and a
no-endorsement clause. What we ship from it is term names --- $6$ class IRIs and $21$ of the
$173$ concepts in its probable-cause scheme --- and not the ontology files, so the clauses
requiring a redistribution to retain the notice do not reach the dataset; the advertising
clause does, because this paper mentions the use, and satisfying it is an acknowledgement
sentence, which the release carries. RFC~8632 and RFC~8345 are Simplified BSD; TMF642 is
Apache-2.0; Topology Zoo is CC~BY~4.0.

\paragraph{Acknowledgement required by NORIA-O's licence.} This product includes software
developed by Orange. That sentence is the acknowledgement NORIA-O's BSD-4-Clause licence
requires of material describing use of it. It is a statement about the ontology and not
about any commercial relationship, and Orange does not endorse this work.

\begin{figure}[htbp]
\centering
\begin{tikzpicture}[
  every node/.style={font=\small},
  ffield/.style={draw, rounded corners=2pt, align=left, inner sep=4pt,
            text width=3.6cm, minimum height=7mm, font=\small\ttfamily},
  fterm/.style={draw, rounded corners=2pt, align=left, inner sep=4pt,
            text width=5.4cm, minimum height=7mm, fill=black!6},
  ar/.style={-{Latex[length=2mm]}, semithick}
]
\node[font=\small\bfseries] at (0,1.0) {emitted field};
\node[font=\small\bfseries] at (7.0,1.0) {term it must resolve to};

\node[ffield] (a1) at (0,0)    {alarm\_type};
\node[ffield] (a2) at (0,-1.1) {probable\_cause};
\node[ffield] (a3) at (0,-2.2) {perceived\_severity};
\node[ffield] (a4) at (0,-3.3) {specific\_problem};
\node[ffield] (a5) at (0,-4.4) {resource, link};
\node[ffield] (a6) at (0,-5.5) {alarm payload};

\node[fterm] (b1) at (7.0,0)    {X.733 event type, via NORIA-O};
\node[fterm] (b2) at (7.0,-1.1) {NORIA-O probable-cause catalogue};
\node[fterm] (b3) at (7.0,-2.2) {RFC 8632 severity};
\node[fterm] (b4) at (7.0,-3.3) {free text --- the only unbound field};
\node[fterm] (b5) at (7.0,-4.4) {RFC 8345 network topology};
\node[fterm] (b6) at (7.0,-5.5) {TMF642 Alarm Management};

\foreach \i in {1,2,3,5,6} \draw[ar] (a\i) -- (b\i);
\draw[ar, dashed] (a4) -- (b4);

\node[draw, thick, rounded corners=2pt, align=center, inner sep=5pt,
      text width=10.5cm, font=\small] (gate) at (3.5,-7.1)
  {A term outside these vocabularies \textbf{fails the run}. The binding is
   checked at generation time, not asserted in prose.};
\end{tikzpicture}
\caption{What each emitted field binds to. NORIA-O references neither RFC 8632
nor TMF642; that binding is ours. \texttt{specific\_problem} is deliberately
free text, and Section~\ref{sec:benchmark} shows why it had to be withheld from
the methods.}
\label{fig:binding}
\end{figure}
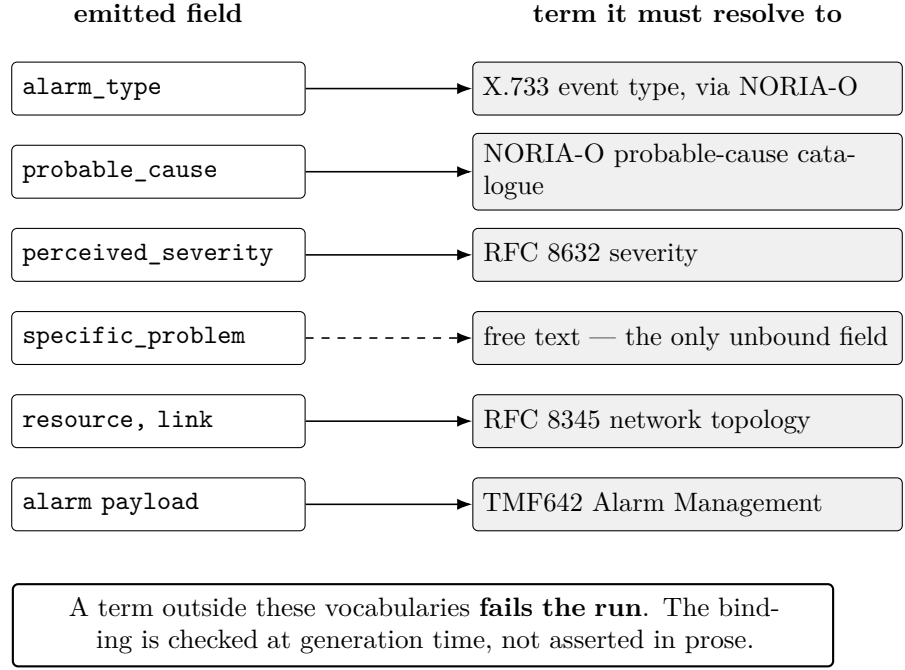

\section{The benchmark}
\label{sec:benchmark}
\paragraph{The task.} One incident at a time, rank the resources that could hold its root
cause, best first. An incident is the NOC's grouping of alarms --- imperfect on purpose,
since some alarms are attached to the wrong one --- so the method is given the storm an
operator would actually see rather than a clean partition.

\paragraph{Scoring.} A ranking is correct at $k$ if one of its first $k$ entries is the
device of any resource the fault was at. Two details matter and both are properties of the
faults rather than choices: a fibre cut has two ends and either is correct, and a site
power loss names the PoP, so any device hosted in it counts. We report $k = 1$ and $k = 3$:
top-1 is what an automated remediation step needs, and top-3 is what an operator who will
check three devices needs. They are different questions and a method can be good at one and
not the other.

An eval incident that raises no alarm anywhere stays in the denominator and is charged a
miss to every method. There are two such incidents in this configuration, and dropping
them --- which both of our scorers did until 2026-10-21 --- reports a rate over 38 trials
under a heading that says 40 (Section~\ref{sec:limitations}).

\paragraph{What a method sees.} The static topology and inventory, and the alarms in the
incident window. Not the fault, not the propagation path, not the repair.

\paragraph{What a method must not see, and how we check.} Ground truth travels with the
data --- \texttt{Fault} nodes, their \texttt{AT} edges to the true root resources, and a
\texttt{CAUSED} edge to each alarm they produced. That is deliberate and labelled, and a
method may not read it. The risk a labelled boundary does not cover is ground truth
reaching an \emph{ordinary} property, where nothing marks it, and this dataset shipped
exactly that: every alarm carried its role in the fault --- root, neighbour, background ---
in the X.733 \texttt{specificProblem} field. Reading that one field scored $0.700$ top-1 on
the dataset as it then stood --- level with the best baseline that dataset had --- while
ranking nothing.

So the boundary is now measured rather than asserted. We score every (property, value) pair
that appears on an alarm in the released files as a one-line method --- rank the devices
whose alarm carries that value, earliest first --- in the benchmark's own units: $112$ pairs
over the $40$ eval incidents. Table~\ref{tab:leak} gives the fields that matter. A property
is \emph{supposed} to correlate with the fault; what must not exist is a value that names
the root almost every time it appears \emph{and} appears in most incidents, which is the
pair of conditions that made the old field free. The role now travels only on
\texttt{CAUSED.via} and in \texttt{ground\_truth.jsonl}, and
\texttt{specificProblem} is a function of \texttt{probableCause} alone, so it cannot carry
side information; a test asserts the two score identically.

\begin{table}[t]
\centering
\begin{tabular}{@{}llrrr@{}}
\toprule
Alarm property & best value & Top-1 & Prec. & Inc. \\
\midrule
\texttt{specific\_problem} \emph{(before)} & \texttt{root} & 0.700 & 1.000 & 28 \\
\midrule
\texttt{perceived\_severity} & \texttt{critical}  & 0.625 & 0.893 & 28 \\
\texttt{specific\_problem} & \texttt{Transmit module failure} & 0.225 & 1.000 & 9 \\
\texttt{probable\_cause} & \texttt{transmitterFailure}  & 0.225 & 1.000 & 9 \\
\texttt{alarm\_type} & \texttt{communicationsAlarm} & 0.225 & 0.250 & 36 \\
\bottomrule
\end{tabular}
\caption{Each alarm field's best single value read as a one-line RCA rule, GÉANT small,
$n=40$. \emph{Prec.} is the share of the incidents where such an alarm exists in which its
earliest one sits on a true root; \emph{Inc.} is how many of the 40 those are. The top row
is the leak, measured with the same tool on the dataset as it was shipped before
2026-10-21, and it is the only row not measured on the released data. The row below it is
not a leak and is the highest on the current dataset, at $0.625$ --- level with the best
method in Table~\ref{tab:results}: \texttt{perceived\_severity} is an RFC 8632 field a real
console shows, and what its score says is that our fault model marks the root critical too
reliably (Section~\ref{sec:limitations}). The two rows below that reach $1.000$ precision
on $9$ incidents each, which the gate deliberately allows: a value that names the root
whenever it appears but appears in a fifth of incidents is a narrow cause doing its job,
not a free answer.}
\label{tab:leak}
\end{table}

\section{Realism study}
\label{sec:realism}

A generator is only worth using if its difficulty is the real task's difficulty. The
reference is the Huawei PCIC 2021 competition data, released under the MIT licence
\cite{gcastle_dataset,qiao2023structural}: subset \texttt{18V\_55N\_Wireless} holds $34{,}838$ timestamped alarms
over 18 alarm types and 55 devices across 240 days, with a ground-truth causal graph over
alarm \emph{types} and no per-incident root-cause labels. It can therefore say whether our
stream is plausible; it cannot score a method.

Four quantities are computable on both sides from $(\text{type}, \text{device},
\text{time})$ alone, and the same definition is applied to both. Bursts split on a 60\,s
quiet gap. Table~\ref{tab:realism} reports them.

\begin{table}[t]
\centering
\begin{tabular}{@{}lrr@{}}
\toprule
 & Generated & Real (PCIC) \\
\midrule
alarms                        & $1{,}572$ & $34{,}838$ \\
alarm types                   & 21    & 18 \\
devices                       & 717   & 55 \\
\midrule
inter-arrival median (s)      & $60.4$  & $132.0$ \\
share of gaps under 1\,s      & $0.041$ & $0.019$ \\
largest burst (alarms)        & $\mathbf{82}$ & $\mathbf{30}$ \\
singleton bursts              & $0.756$ & $0.712$ \\
\midrule
share of the commonest type   & $0.167$ & $0.212$ \\
type entropy, normalised      & $0.784$ & $0.839$ \\
\midrule
excitation lift, maximum      & $524.0$ & $69.48$ \\
pairs at lift $\geq 2$, share & $0.150$ & $0.145$ \\
\bottomrule
\end{tabular}
\caption{Generated GÉANT small against real operator alarms. The type field is
\texttt{probable\_cause}, X.733's catalogue of named causes, which is the analogue of
PCIC's \texttt{alarm\_id}; see the caveat below on the field this table used first.
Excitation is a lift-over-base-rate \emph{proxy}: for each ordered type pair, how often
$B$ follows $A$ within 300\,s divided by $B$'s base rate. No Hawkes kernel is fitted and
no parameter is estimated on either side; a fit remains unmeasured. Bold marks the one
axis on which the two streams differ.}
\label{tab:realism}
\end{table}

\paragraph{One divergence left, and the one we fixed changed the results.} The first
version of this study found the generated stream firing in near-simultaneous volleys:
$0.205$ of consecutive alarms under a second apart against $0.019$ of real ones. The cause
was one line of the fault model. Every neighbour-cell alarm in a radio storm was drawn from
a single lognormal around the fault's onset, so a dead site produced a cloud rather than a
sequence: $283$ of the $332$ sub-second gaps in the stream were neighbour alarms. A live
site does not notice a dead neighbour the moment it dies --- it detects the degradation on
its own measurement period --- so each site now draws an offset into a 15-minute period.
The sub-second share fell to $0.041$, the median inter-arrival rose from $40.6$\,s to
$60.4$\,s against $132.0$\,s real, and the singleton-burst share moved from $0.824$ to
$0.756$ against $0.712$ real. Section~\ref{sec:results} is measured after that fix, and
Section~\ref{sec:limitations} records what it moved.

What remains is burst size: our largest burst is $82$ alarms against $30$. We tested the
obvious explanation --- that a bigger network gives one incident more elements to affect ---
and it is wrong. Across the sweep of Section~\ref{sec:sweep}, Abilene at $144$ devices has
a larger mean largest-burst than Cogentco at $2{,}625$, so burst size here is set by the
fault model and the alarm rates and not by the size of the graph. We have no mechanism for
the remaining factor of $2.7$ and do not claim one.

On the other two axes the generated stream matches. The type vocabulary is 21 named causes
against 18, slightly flatter than the real one rather than more concentrated (normalised
entropy $0.784$ against $0.839$; commonest type $0.167$ against $0.212$). The share of
ordered type pairs exciting each other above $2\times$ is $0.150$ against $0.145$, a ratio
of $1.03$. The maximum lift does not match and is not comparable: ours is $524.0\times$
against $69.48\times$, because spreading the storms left some ordered pairs rare enough
that a handful of occurrences divides by a very small base rate. A maximum over 441 pairs
is a tail statistic on a 1,572-alarm stream; the share above a threshold is the one to
read, and it is the one we read.

\paragraph{The field this table measures is a choice, and the first version chose wrong.}
The analogue of PCIC's \texttt{alarm\_id} --- a named vendor alarm --- is X.733's
\texttt{probable\_cause}, of which this dataset uses 21 values. The first version of this
study used \texttt{alarm\_type}, X.733's five-value \emph{event type} enum, and so compared
a severity-class taxonomy against a vendor alarm name. On that field the same alarms at the
same times read 4 types against 18, a commonest-type share of $0.503$ against $0.212$, and
a strongest excitation of $1.95\times$ against $69.5\times$ --- from which we concluded that
our vocabulary was too small and that our alarm types did not chain at all. Both
conclusions were artefacts of the field. With four types the commonest has base rate
$0.51$, so a lift of $2\times$ is near the arithmetic ceiling and the measure had almost no
room to move. Same data, different label column, opposite finding. The results file carries
both fields so the choice is auditable rather than buried.

\paragraph{The test that matters: the ranking mostly transfers.} Whether the \emph{order}
of methods on generated data holds on real data decides whether the generator is useful for
method development. The four topology-free rules, scored on $4{,}539$ real alarm bursts and
on our 40 eval incidents, agree on the top two and disagree on the bottom two: earliest
then most-frequent on both sides, then real data puts random above most-recent and ours
does not, for a Kendall $\tau$ of $0.667$. The two it swaps are tied on our side at $0.225$
each, so the disagreement is between two rows our own data cannot order. An earlier version
of this study read $\tau = 1.0$; that was on the generated stream before the arrival-time
fix of Section~\ref{sec:realism}, and we report the weaker number measured after it.

What does transfer is the finding that matters. The earliest alarm names a root cause in
$0.773$ of real bursts against $0.625$ of ours, so the rule that beats our methods is not
an artefact of the generator; it is how operator alarm streams behave.

\paragraph{Four caveats, all load-bearing.} The per-burst label on the real side is
\emph{derived by us} from the dataset's type-level causal graph --- a type is a root if no
other type present in the burst is its ancestor --- because PCIC ships no per-incident
labels; that definition encodes the same ``a cause precedes its effects'' assumption our
generator does, and can flatter a rule that shares it. The real task is type-level and
ours is device-level, so only the ordering is comparable, never the rates. Our own two
graph methods need a propagation graph built from device topology, which
\texttt{18V\_55N} does not carry, so they are absent from the comparison: the transfer
claim is about the baselines' order, not our margin over them. And four rules admit 24
orderings, so an exact match carries $p = 0.042$ under a uniform null --- one ranking of
four things, not a strong test. Of the $21{,}427$ real bursts, $16{,}071$ contain a single
alarm type and 817 have no causal edge among the types present; both classes make every
rule identical, and the $4{,}539$ that remain are the comparison. That skip is a property
of the real stream, not a filter chosen to help a rule.

\paragraph{The second PCIC subset is labelled and still unusable.} \texttt{espk} carries
$141{,}853$ alarms, a 35-device topology, and 49 incidents labelled
(alarmed device set $\rightarrow$ root-cause device) --- device-level and labelled, which
is exactly our task. Its snapshots carry no timestamps, so no time-ordering method can run
on them, and its 49 incidents share only 4 distinct root causes over a median snapshot of
28 of 35 devices, which a constant predictor exploits. We report it rather than use it.

\section{Baselines}
\label{sec:baselines}
Nine rows are scored. Four are one-line rules, two are the graph methods we designed, and
three are baselines we did not design --- the last group matters because a resource paper
whose own method is the best row has not shown the resource is useful to anyone else.

\paragraph{Rules.} Earliest critical alarm; most-alarmed device; most recent change within
two hours of the first alarm; and a uniform draw among the alarmed devices, which is the
floor.

\paragraph{Ours.} \emph{temporal} ranks candidates by how many symptoms they explain under
a backwards walk over edges derived from topology and alarm times, breaking ties by onset.
\emph{neighbourhood} is the same candidate set with time removed, ranking by how many
alarmed devices a candidate neighbours; it is in the table to show what the time ordering
contributes.

Both run on our own graph engine, Samyama, through its \texttt{algo.symptom\-Explanation}
procedure: the walk and the explanation count are the engine's, and what we wrote is the
construction of the edges it walks. The runs reported here used the embedded Python SDK
\textbf{1.9.0}. We name the version because these two rows cannot be reproduced without
it, and because the engine is ours --- a resource paper that leaves its own dependency
unversioned is asking the reader to take the rows on trust. A later engine release would
be a different measurement and we have not made it.

\paragraph{External.} \emph{thp\_causal} is gCastle's \texttt{TTPM}
\cite{gcastle_dataset}, the reference implementation of the topological Hawkes process of
Cai et al.\ \cite{qiao2023structural}, under Apache-2.0. It is fitted once, on the 697
alarms raised before the first eval incident begins, and sees no eval alarm, no incident
grouping and no ground truth; it is given the same device adjacency the other methods get.
It learns 41 edges over the 21 \texttt{probable\_cause} values. That graph is type-level,
so a reduction is needed to answer a device question, and we use the published one: among
the cause types present in an incident, keep those with no parent also present, and rank
the devices that raised one of them, earliest first.

\emph{thp\_nograph} is that same reduction with the learned graph removed, so every cause
type present counts as root --- which makes it ``the device that alarmed earliest''. It is
the control. Without it the \emph{thp\_causal} row cannot be read at all, because a change
could equally be the causal model or the reduction wrapped around it.

\emph{pagerank\_alarmed} is personalised PageRank over the same device adjacency,
teleporting uniformly to the alarmed devices --- the MonitorRank and CloudRanger shape. It
uses no time and no causal model, ranking purely by structural centrality within the
alarmed set. It cannot name a PoP, so a site-wide cause is reachable for it only through
the scoring rule that a device at the root PoP counts as a hit.

\paragraph{An LLM, given the same view.} \emph{llm\_ranker} sends one prompt per incident
to a general-purpose model (claude-sonnet-5-5) containing exactly what
Section~\ref{sec:benchmark} allows: the incident's alarms with their times relative to the
first, the role of each alarmed device, the adjacency around them, and the candidate list.
It asks for three candidate ids, best first. It is the only row whose input is free text,
so what reaches it cannot be audited by reading a query; a test asserts over the generated
prompt that no fault id, scenario id, mechanism name or absolute timestamp appears in it.

Two of its properties are unlike every other row and both are recorded with the number. It
is not deterministic and this is one run, so $0.600$ is a sample from a distribution rather
than a repeatable figure. And a reply can be unusable: an answer that names no candidate is
charged a miss, never retried, because a method that sometimes returns nothing scores lower
and a retry loop would measure a different method. The run cost \$$2.30$, which makes it and
the agent below the only baselines here with a price per query.

\paragraph{The same model, allowed to ask.} \emph{llm\_agent} is the agent row. It gets ten
read-only verbs over the graph --- an incident's alarms, a node's properties, a device's
cable links, a path between two devices, the devices at a site, the changes targeting a
device, the services depending on it, how many alarms it raised before this incident ---
and $6$ rounds of queries before it must answer. It has no shell and no file access: it
emits query lines, we execute them, and the verb set is closed, so a request we do not
implement is answered with the verb list.

The graph it queries is sanitised before it sees anything: \texttt{Fault} nodes, \texttt{AT}
and \texttt{CAUSED} edges are removed and the \texttt{scenario} field is stripped from every
alarm and incident. That last one matters more than it looks --- an incident's id is
\texttt{inc:flt:eval-overlap-0050:0}, which names its own fault, and the scenario names the
split and the index. A ranker is safe because its prompt is a fixed string a test can read;
an agent chooses its own queries, so the only durable guarantee is that the answer is not in
the graph. We assert that in a test rather than trusting the verb list.

\section{Results}
\label{sec:results}

\begin{figure}[htbp]
\centering
\includegraphics[width=0.78\linewidth]{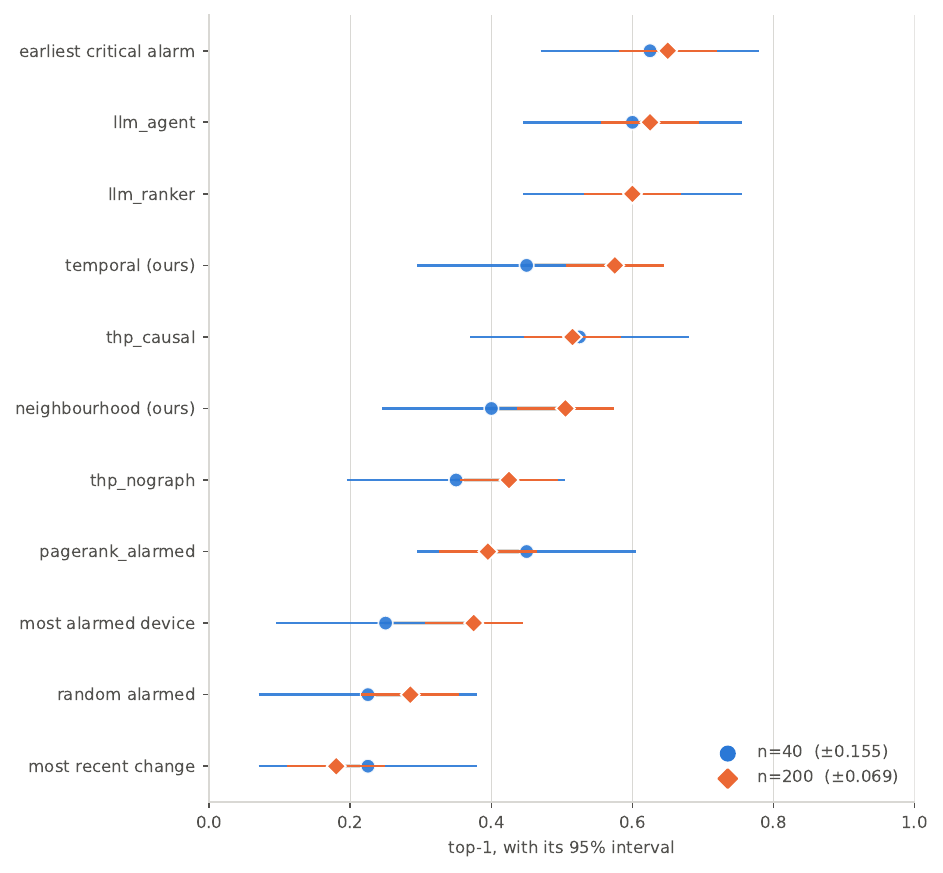}
\caption{Every method at both sample sizes, with its 95\% interval. The $n=40$ intervals
(circles) are more than twice the width of the $n=200$ ones (diamonds), and most of the
table sits inside a single one of them.}
\label{fig:samplesize}
\end{figure}

\paragraph{Forty incidents was not enough, and we can say exactly how.} Table~\ref{tab:results}
reports every method at two sample sizes. The $n=40$ column is the configuration this study
first used; the $n=200$ column is five times as many incidents from the same generator at the
same seed with the same rates, so the two differ in how much was measured and in nothing
else. Figure~\ref{fig:samplesize} draws both with the interval each carries.

The smaller sample was not merely imprecise, it was wrong about the order. Three rows move
by $0.1$ or more --- our temporal method by $0.125$, our neighbourhood method by $0.105$,
the most-alarmed rule by $0.125$ --- and the ordering changes with them: at $n=40$ our
temporal method sits below \emph{thp\_causal}, and at $n=200$ it is above it. Forty
incidents can separate the top of this table from the bottom and cannot order the middle.
We report the larger column as the result and keep the smaller one visible, because a
reader deserves to see what a $5\times$ smaller sample of the same data said.

\begin{table}[t]
\centering
\begin{tabular}{@{}lrrr@{}}
\toprule
& \multicolumn{1}{c}{$n=40$} & \multicolumn{2}{c}{$n=200$} \\
\cmidrule(lr){2-2}\cmidrule(lr){3-4}
Method & Top-1 & Top-1 & Top-3 \\
\midrule
\multicolumn{4}{@{}l}{\itshape rules we did not design} \\
earliest critical alarm & $0.625$ & \textbf{0.650} & $0.650$ \\
most alarmed device & $0.250$ & $0.375$ & $0.535$ \\
random among alarmed devices & $0.225$ & $0.285$ & $0.545$ \\
most recent change & $0.225$ & $0.180$ & $0.200$ \\
\midrule
\multicolumn{4}{@{}l}{\itshape external, not designed by us} \\
llm\_ranker (claude-sonnet-5-5)$^\dagger$ & $0.600$ & $0.600$ & $0.670$ \\
llm\_agent (same model, 10 graph verbs)$^\dagger$ & $0.600$ & $0.625$ & $0.670$ \\
thp\_causal (gCastle TTPM) & $0.525$ & $0.515$ & $0.655$ \\
thp\_nograph \emph{(control)} & $0.350$ & $0.425$ & $0.625$ \\
pagerank\_alarmed & $0.450$ & $0.395$ & $0.600$ \\
\midrule
\multicolumn{4}{@{}l}{\itshape ours} \\
neighbourhood, no time & $0.400$ & $0.505$ & $0.670$ \\
temporal & $0.450$ & $0.575$ & \textbf{0.790} \\
\bottomrule
\end{tabular}
\caption{GÉANT, the same generator and seed at two sample sizes: $40$ eval incidents
and $200$. The configurations differ in scenario counts and in nothing else. Two of them raise no alarm on any device,
so no method has anything to rank; each is charged a miss to every row rather than dropped
(Section~\ref{sec:limitations}). A rate near $0.5$ carries $\pm 0.155$ at 95\% on $40$ trials and $\pm 0.069$ on $200$;
Figure~\ref{fig:samplesize} draws both. No paired test has been run.
$^\dagger$ one run of a non-deterministic method, so that row is a sample and not a
repeatable number; the others are. Every row is measured after the arrival-time fix of
Section~\ref{sec:realism}; the numbers in the first version of this paper were measured
before it and all of them moved.}
\label{tab:results}
\end{table}

\paragraph{The simplest rule usually wins at top-1.} Ranking the device that raised the
earliest critical alarm takes $0.625$, against $0.450$ for our temporal method and $0.525$
for \emph{thp\_causal}, the best non-LLM baseline. Across the $15$ runs of
Section~\ref{sec:sweep} it takes top-1 in $13$ of them, so ``usually'' is the strongest
form of this claim we can make: in the other $2$ our temporal method does. Its margin is
narrower than it was: on the pre-fix stream it took $0.700$ against $0.475$, and spreading
the storms cost the rule $0.075$ while costing our temporal method $0.025$. A rule that
reads the front of the queue loses more when the queue stops arriving all at once. Section~\ref{sec:realism} tested whether this is an artefact of
the generator, and it is not: the same rule takes $0.773$ on real operator alarms, higher
than on ours. Part of the margin is still ours to answer for --- our storms are
$10\times$ more likely to arrive inside one second (Table~\ref{tab:realism}), which
shortens the queue the rule has to pick from --- but the finding survives the generator.
Tuning the method against this number would be fitting to our own data and is not done.

\paragraph{The word doing the work is ``critical'', not ``earliest''.}
\emph{thp\_nograph} is the earliest-alarm rule without the severity filter, and it takes
$0.350$ against the earliest \emph{critical} rule's $0.625$. The two differ in nothing else:
same candidate set, same scoring function, same 40 incidents. The gap widened when the
storms were spread: on the pre-fix stream the control took $0.475$ against the rule's
$0.700$, so dropping the severity filter costs more, not less, once alarms stop arriving in
one volley. So the one-line rule's strength is not that it reads the clock; it is that the
generator marks the root alarm critical often enough for a severity filter to find it. That
is a property of our fault model and the sharpest single thing to attack in the next
version of the generator.

\paragraph{The learned causal graph helps, slightly, and only at top-1.}
\emph{thp\_causal} takes $0.525$ against its own control's $0.350$: restricting the
earliest-first ranking to causally-root cause types promotes the right device in seven more
incidents out of 40, and it is now the best method in the table that is neither a one-line
rule nor an LLM. On the pre-fix stream the same comparison was $0.525$ against $0.475$, so
the whole of this gain appeared when the arrival times were fixed: the control degrades
without the volley and the causal filter does not. At top-3 the two are level at $0.625$.
The top-1 difference is seven incidents wide against a $\pm 0.155$ interval, which is the
first result here where the learned structure does something a single draw can see --- and
it is still one draw, from a kernel fitted on 688 alarms.

\paragraph{A model that knows what an alarm means still loses to the severity field.}
\emph{llm\_ranker} takes $0.600$ top-1 --- above every graph method in the table, above
both of ours, and below the $0.625$ of a rule that reads one field. Whatever telecom
knowledge the model brings is worth something over structure-only ranking and is not worth
as much as knowing that the generator marks root alarms critical. Its top-3 is the more
interesting number: $0.625$, barely above its own top-1, where our temporal method rises
from $0.450$ to $0.725$ given the same two extra slots. Of 40 incidents it is right first
in 24, second in 1, and third in none. It returns one confident answer and a tail it has
not really ranked, which is exactly the shape that does not help an operator who will check
three devices.

Its per-fault-type scores say where the knowledge pays, and one class where it buys
nothing at all: $0.000$ on every site power loss, at both sample sizes. That class is the
subject of the next paragraph, and it is the one place in this paper where the methods we
designed are not merely competitive but alone.

\paragraph{Giving the model tools changed what it looked at and not what it answered.}
\emph{llm\_agent} is the same model with ten read-only graph verbs and six rounds of
queries before it must answer: it can list an incident's alarms, walk cable links, find a
path between two devices, pull the change records near a device, and ask how many alarms a
device raised before this incident. It used them --- $263$ queries, $6.58$ per incident,
never exhausting its turns --- and scored $0.600$ top-1 and $0.625$ top-3, identical to the
ranker that saw one fixed prompt, at $2.7\times$ the cost (\$$6.14$ against \$$2.30$). Its
per-fault-type scores are identical too, in all six classes.

The two agree on the first pick in $38$ of $40$ incidents; they differ only in the second
and third slots, which this benchmark's scoring barely rewards. What the agent chose to
look at is the more useful result: $101$ of its queries asked for change records and $29$
for a device's alarm history, against $66$ for the topology and $35$ for the incident's own
alarms. It spent most of its budget on the hypothesis that a recent change caused the
fault --- which is the worst-scoring baseline in Table~\ref{tab:results} at $0.225$ --- and
on $7$ incidents it answered without issuing a single query. We read this as evidence that
the task is not tool-limited: the information needed to beat $0.625$ is in the alarm list,
and no amount of graph querying supplies what is missing.

\paragraph{When the root is silent, only a method that reads the topology works.} A site
power loss kills the PoP, and a dead PoP raises no alarm. Every alarm in the storm that
follows belongs to a device that is working, so a method that \emph{ranks alarmed devices}
cannot name the root --- not because it ranks them badly, but because the answer is not in
the list it ranks. The $12$ power-loss incidents at $n=200$ say exactly that:

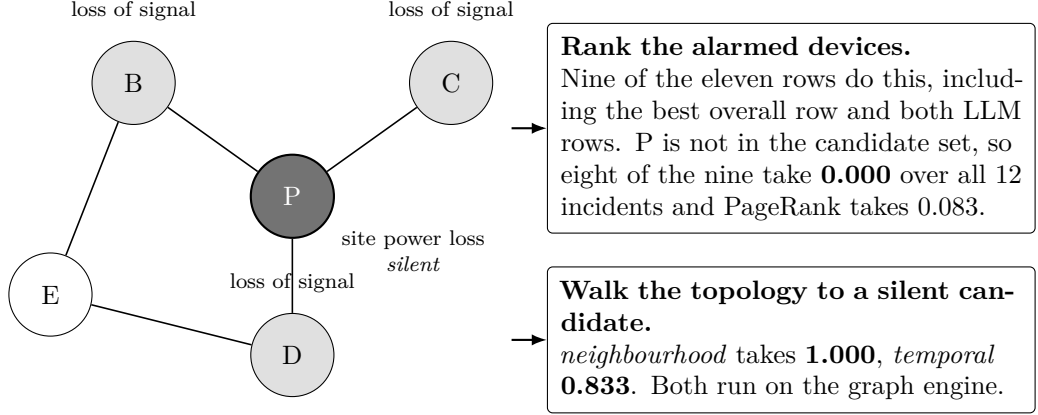
\begin{figure}[htbp]
\centering
\begin{tikzpicture}[
  every node/.style={font=\small},
  pop/.style={draw, circle, minimum size=11mm, inner sep=1pt, font=\small},
  dead/.style={pop, fill=black!55, text=white, thick},
  alarmed/.style={pop, fill=black!12},
  lk/.style={semithick},
  ar/.style={-{Latex[length=2mm]}, thick},
  note/.style={draw, rounded corners=2pt, align=left, inner sep=5pt,
               text width=6.1cm, font=\small}
]
\node[dead]    (p)  at (0,0)      {P};
\node[alarmed] (b)  at (-2.1,1.5) {B};
\node[alarmed] (c)  at (2.1,1.5)  {C};
\node[alarmed] (d)  at (0,-2.1)   {D};
\node[pop]     (e)  at (-3.2,-1.3){E};

\draw[lk] (p) -- (b); \draw[lk] (p) -- (c); \draw[lk] (p) -- (d);
\draw[lk] (b) -- (e); \draw[lk] (e) -- (d);

\node[font=\scriptsize, align=center] at (1.6,-0.72) {site power loss\\\emph{silent}};
\foreach \n in {b,c,d}
  \node[font=\scriptsize] at ($(\n)+(0,0.95)$) {loss of signal};

\node[note] (bad) at (6.6,0.9)
  {\textbf{Rank the alarmed devices.}\\
   Nine of the eleven rows do this, including the best overall row and both
   LLM rows. P is not in the candidate set, so eight of the nine take
   \textbf{0.000} over all $12$ incidents and PageRank takes $0.083$.};
\node[note] (good) at (6.6,-1.9)
  {\textbf{Walk the topology to a silent candidate.}\\
   \emph{neighbourhood} takes \textbf{1.000}, \emph{temporal}
   \textbf{0.833}. Both run on the graph engine.};

\draw[ar] (2.9,0.9) -- (bad);
\draw[ar] (2.9,-1.9) -- (good);
\end{tikzpicture}
\caption{The fault class that separates the table. A dead PoP cannot report
itself, so the answer is the one device that emitted nothing. Reaching it is a
traversal over the topology, and no amount of reading the alarm stream
substitutes for it.}
\label{fig:powerloss}
\end{figure}

\begin{center}
\begin{tabular}{@{}lr@{}}
\toprule
On the $12$ site power losses & Top-1 \\
\midrule
neighbourhood (ours) & \textbf{1.000} \\
temporal (ours) & $0.833$ \\
pagerank\_alarmed & $0.083$ \\
earliest critical alarm & $0.000$ \\
most alarmed device & $0.000$ \\
thp\_causal & $0.000$ \\
llm\_ranker & $0.000$ \\
llm\_agent & $0.000$ \\
\bottomrule
\end{tabular}
\end{center}

Our two methods nominate candidates that never alarmed, by walking the topology from the
devices that did: a PoP whose hosted devices all went quiet at once is itself a candidate.
Everything else in the table --- four rules, a fitted topological Hawkes process,
personalised PageRank, a language model with the alarms in front of it, and the same model
allowed ten graph queries --- scores zero or near it. This is not a tuning difference; it is
the difference between selecting from a list and inferring something absent from it.

\begin{figure}[htbp]
\centering
\includegraphics[width=0.95\linewidth]{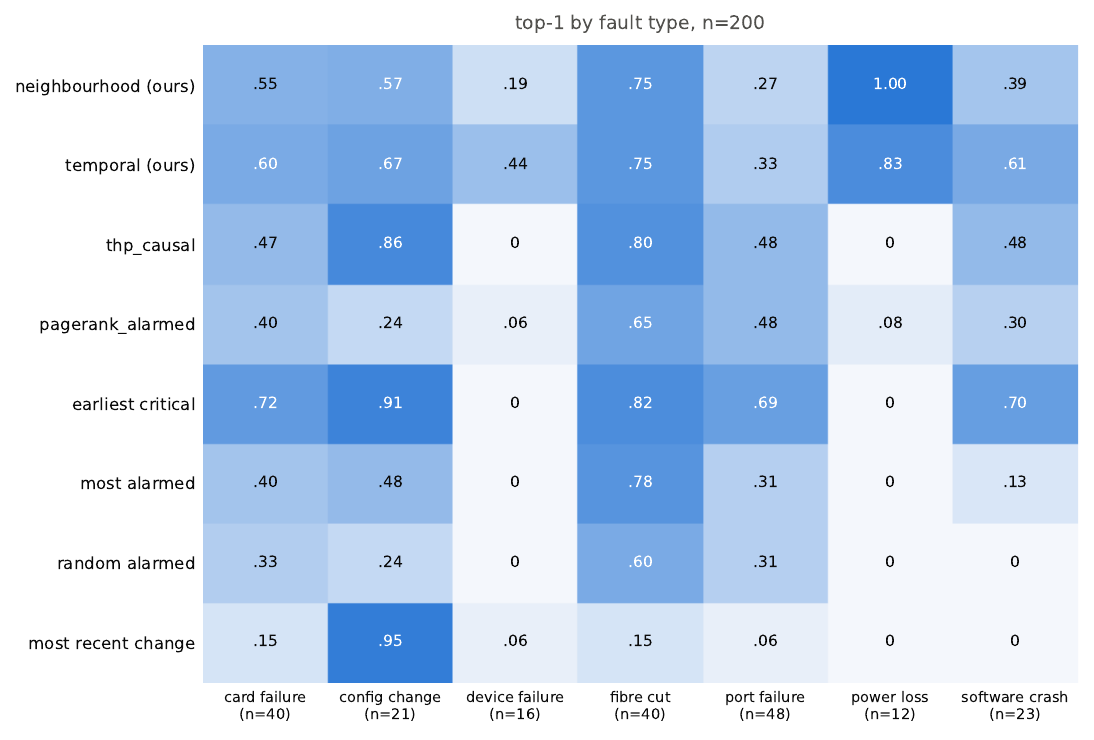}
\caption{Top-1 per fault type at $n=200$. Reading down the \emph{power loss} column: the
two methods that walk the topology take $1.000$ and $0.833$, and every rule, both Hawkes
variants and both LLM rows take $0.000$. No other column separates the table that way.}
\label{fig:byfaulttype}
\end{figure}

It is also the result that $40$ incidents could not show. At $n=40$ this class held $3$
incidents, the paper reported that no method did well on it, and that statement was wrong:
it was three samples of a class our methods solve and nothing else touches.
Figure~\ref{fig:byfaulttype} shows the whole grid, where the power-loss column is the only
one with a clean split.

\paragraph{Where else the methods win.} Our temporal method takes top-3 at $0.790$, the
best in the table --- the next row is $0.670$ --- and the earliest-critical rule cannot improve on its own
top-1 at all --- $0.650$ at both $k=1$ and $k=3$, because it returns one candidate. That is a comparison
between a ranker and a classifier, and the gain should be read as such: a top-3 list is
worth something to an operator who will check three devices, and nothing to an automated
remediation step that acts on one. We claim the former only.

\subsection{What the graph engine contributes, and what it does not}
\label{sec:engine-value}
The \emph{neighbourhood} and \emph{temporal} rows are the only two that run on a graph
database, and Figure~\ref{fig:powerloss} is the only place in the grid where that is
decisive. The reason is structural rather than incidental: every other row consumes the
alarm stream and returns something from it, so its answer is bounded by the set of devices
that spoke. A traversal is not. \texttt{algo.symptomExplanation} walks outwards from the
alarmed set over edges we construct from topology and alarm times, and a node it reaches
is a candidate whether or not it emitted anything --- which is why a PoP whose hosted
devices all went quiet at once can be nominated at all. That is one capability, stated
once: the ability to name a vertex that is absent from the evidence.

We claim nothing about speed or scale here, and the reader should not infer it. The
graph these two rows run on is $11{,}659$ nodes and $14{,}793$ edges, which fits
anywhere --- and the largest backbone in Section~\ref{sec:sweep} is bigger only in
proportion to its device count, which is $2{,}625$ against $717$; no
query latency, memory figure or throughput comparison appears in this paper, and none
should be read into the two rows. The engine is reported as a dependency with a version,
not as a result. Equally, the capability is not unique to our engine --- any system that
can traverse the topology can express the same walk, and we would expect a reimplementation
to land on the same two numbers. What Figure~\ref{fig:powerloss} measures is that the
traversal is worth doing, not that ours is the only way to do it.

\subsection{Across seeds and topologies}
\label{sec:sweep}

We re-ran the whole benchmark $15$ times: three Internet Topology Zoo backbones
---~GÉANT ($40$ PoPs), Abilene ($11$) and Cogentco ($197$)~--- at five seeds each, one
config throughout, scored by the same script. A seed redraws the network's failure history
\emph{and} the $40$ eval incidents, so each run is a different benchmark instance of the
same generator.

\begin{table}[t]
\centering
\footnotesize
\setlength{\tabcolsep}{4pt}
\begin{tabular}{@{}lcccccc@{}}
\toprule
& \multicolumn{2}{c}{GÉANT ($40$)} & \multicolumn{2}{c}{Abilene ($11$)}
& \multicolumn{2}{c}{Cogentco ($197$)} \\
\cmidrule(lr){2-3}\cmidrule(lr){4-5}\cmidrule(lr){6-7}
Method & mean & range & mean & range & mean & range \\
\midrule
earliest critical alarm      & $0.585$ ($\pm0.038$) & $0.550$--$0.625$ & $0.620$ ($\pm0.048$) & $0.575$--$0.700$ & $0.595$ ($\pm0.054$) & $0.525$--$0.650$ \\
temporal (ours)              & $0.485$ ($\pm0.055$) & $0.450$--$0.575$ & $0.565$ ($\pm0.076$) & $0.475$--$0.650$ & $0.535$ ($\pm0.029$) & $0.500$--$0.575$ \\
neighbourhood (ours)         & $0.420$ ($\pm0.062$) & $0.375$--$0.525$ & $0.480$ ($\pm0.084$) & $0.350$--$0.550$ & $0.485$ ($\pm0.065$) & $0.400$--$0.575$ \\
thp\_causal                  & $0.395$ ($\pm0.116$) & $0.225$--$0.525$ & $0.420$ ($\pm0.076$) & $0.350$--$0.525$ & $0.380$ ($\pm0.021$) & $0.350$--$0.400$ \\
\bottomrule
\end{tabular}
\caption{Top-1 over five seeds per topology, mean ($\pm$ sd) and observed range; $n=40$ in
every run. The number in each heading is the backbone's PoP count. \emph{llm\_ranker} is
absent: it is a paid, non-deterministic method and was run once.}
\label{tab:sweep}
\end{table}

\paragraph{The rule wins most runs, not every run.} The earliest-critical rule takes top-1
in $13$ of the $15$ runs. In the other $2$ our temporal method does, and both are the same
seed on two different backbones. No method wins everywhere, and the full ordering of
Table~\ref{tab:results} reproduces in $1$ of the $15$ runs --- the shipped seed itself.
Every other run permutes the middle of the table somewhere. The headline claim survives
in the form ``usually'' and does not survive in the form ``always''.

\paragraph{The seed spread was itself an artefact of the arrival times.} At $n=40$ one
run's rate carries a standard error of $0.079$, and $\pm 0.155$ at 95\% is $1.96$ of those.
The seed-to-seed sd of the best row is now $0.038$ on GÉANT, $0.048$ on Abilene and $0.054$
on Cogentco --- that is $0.48$, $0.61$ and $0.68$ of one run's standard error, so redrawing
the whole network and all 40 incidents moves the rate \emph{less} than resampling 40
incidents would. The $\pm 0.155$ interval is the conservative one to quote.

That is the opposite of what the same sweep said before the arrival-time fix of
Section~\ref{sec:realism}, and the reversal is the finding. On the pre-fix stream the GÉANT
sd was $0.127$, or $1.61$ standard errors, and the rule's range across five seeds was
$0.300$ wide. Spreading the storms cut that spread to a quarter. The instability was not
seed sensitivity in any interesting sense: it was that a rule reading the front of a
simultaneous volley is at the mercy of which alarm happened to land first, and a few
hundred milliseconds of draw decided whole incidents. Fixing the mechanism removed most of
the variance along with some of the accuracy.

\paragraph{What does hold across all three backbones.} The earliest-critical rule has the
highest mean on all three; our temporal method is second on all three; and
\emph{thp\_causal}, the only method given the topology to learn from, is last of these four
on all three, with the widest spread of any row on GÉANT ($\pm0.116$) and the narrowest on
Cogentco ($\pm0.021$). Network size moves the absolute rates very little: Cogentco has
$197$ PoPs against Abilene's $11$, and the best mean differs by $0.025$ between them. The
generator's difficulty travels with the fault model and the noise process, not with the
graph --- which is also why the remaining burst-size divergence in
Section~\ref{sec:realism} cannot be explained by network size.

\subsection{Which noise source makes it hard}
\label{sec:ablation}

The generator asserts five sources of difficulty, each a config knob. An ablation turns
exactly one off, on the same topology and the same seed, and re-scores: what moves is
attributable to that knob, and what does not move is difficulty we claimed and did not add.
Six runs, the shipped config plus five.

\begin{table}[t]
\centering
\small
\begin{tabular}{@{}lrrrr@{}}
\toprule
Source removed & earliest crit. & temporal & neighbourhood & thp\_causal \\
\midrule
alarms that never arrive & $+0.150$ & $+0.150$ & $+0.200$ & $-0.025$ \\
background alarms$^\ddagger$ & $+0.000$ & $+0.025$ & $+0.075$ & $+0.075$ \\
per-device clock skew & $+0.000$ & $+0.100$ & $+0.075$ & $-0.075$ \\
neighbour-cell storms & $+0.000$ & $+0.100$ & $+0.175$ & $-0.100$ \\
mis-grouped incidents & $+0.050$ & $+0.025$ & $+0.050$ & $+0.075$ \\
\bottomrule
\end{tabular}
\caption{Signed change in top-1 when one noise source is removed; positive means the method
got better without it. GÉANT small, seed as in Table~\ref{tab:results}, $n=40$ throughout.
$^\ddagger$ removing background alarms also changes how many incidents have any alarm to
rank, from $0$ to $3$: for three incidents the only alarms the NOC had attached were
background ones. Three forced misses then sit inside every method's delta on that row, so
it is not a difference in difficulty and we do not read it as one.}
\label{tab:ablation}
\end{table}

\paragraph{Missing alarms are the difficulty; the severity rule is immune to the rest.}
Delivering every alarm that could be raised is the largest move in the table: $+0.150$ for
the earliest-critical rule, $+0.150$ for our temporal method and $+0.200$ for our
neighbourhood method. Alarms that never arrive are what the benchmark is mostly made of.
Three of the other four rows move the earliest-critical rule by exactly $0.000$: neither
clock skew, nor the neighbour storms, nor the background noise changes which device raised
the earliest critical alarm. That is the clearest statement of what the one-line rule is
doing, and of why the noise we add does not trouble it.

Clock skew and the neighbour storms cost the two methods that read order and count
neighbours: $+0.100$ each to our temporal method, $+0.075$ and $+0.175$ to our
neighbourhood method. Both \emph{hurt} \emph{thp\_causal} when removed ($-0.075$ and
$-0.100$), which we do not have a mechanism for and do not read as a result at two or four
incidents wide. Mis-grouping, the source an earlier version of this study found worthless,
now moves four methods by $0.025$ to $0.075$: it was worth nothing while the storms arrived
in a volley and is worth something now that they do not.

\paragraph{One row is not a difference, and a check that watched the wrong thing said it
was.} Removing background alarms takes the count of incidents with no alarm to rank from
$0$ to $3$, because for three incidents every alarm the NOC had attached was a background
one. Three forced misses sit inside that row's deltas. The first version of this check
watched the denominator $n$, which cannot move --- an unrankable incident is charged a miss,
not dropped --- and so reported all five ablations as clean subtractions, including the two
largest moves in the table at the time.

\section{Limitations}
\label{sec:limitations}

\paragraph{An incident with no alarm is charged as a miss, not dropped.} A fault whose
every symptom went unreported raises no alarm on any device, and no method has anything to
rank. Both of our scoring scripts dropped such an incident until 2026-10-21, which scored
38 trials while the split size printed in the caption was 40. An invisible fault is still a
fault the operator did not localise, and a benchmark that deletes what nothing can answer
reports a denominator it does not have, so each is now a miss for every method and every
baseline. The dataset in Table~\ref{tab:results} happens to have $0$ of them, against $2$
before the arrival-time regeneration, so the rule costs nothing here --- it is stated
because it is the rule, not because it binds. The ablation in Table~\ref{tab:ablation}
shows why it matters: switching off background alarms takes that count to $3$.

\paragraph{Nearly half the alarms are background.} $0.471$ of alarms are unrelated to any
injected fault. A method scoring above the random-among-alarmed floor of $0.225$ is
therefore not reading a clean signal --- but neither is the task as hard as a real console,
where background rates vary by time of day and our noise process is stationary.

\paragraph{One configuration, and one seed per row of the main table.} Every number in
Table~\ref{tab:results} is GÉANT small at one seed. Section~\ref{sec:sweep} measures what
that costs and the answer is less than we expected: the best row's seed-to-seed sd is
$0.038$ on that topology, under half of one run's standard error, so the $\pm 0.155$
sampling interval already covers it. What does not reproduce is the \emph{ordering}: the
full order of Table~\ref{tab:results} appears in $1$ of the $15$ runs, because the methods
sit closer together than the spread of any one of them. Read the table as one instance of
the rates and as a weak claim about their order, with Table~\ref{tab:sweep} as the error
bar on both. Three backbones and five seeds are still one config: the fault rates, the
noise rates and the split policy are fixed throughout, and nothing here says what happens
when those change.

\paragraph{We built the data and two of the methods.} The resource is the claim, and the
methods we designed are not the best rows in Table~\ref{tab:results}: a one-line rule beats
both at top-1. Five baselines we did not design are now scored --- three graph or statistical
methods, an LLM ranker and an LLM agent that queries the graph itself --- and none of them
beats that rule either, which is some evidence the task is as it looks rather than tuned to
us. The strongest remaining objection is not a missing baseline but the fault model: the
rule works because our generator marks root alarms critical, and no baseline can tell us
whether a real network does so as often.

\paragraph{Our storms were too bursty, and fixing it moved every number in this paper.}
The version of this study before 2026-10-21 reported $0.205$ of consecutive alarms arriving
within a second against $0.019$ of real ones, and said that fixing the fault model's
emission timing would invalidate every number in its results table. It did, and this is
that paper. The mechanism was one line --- every neighbour-cell alarm in a radio storm drawn
from a single lognormal around the fault's onset --- and the fix moved the sub-second share
to $0.041$. What it cost is recorded throughout: the severity rule fell from $0.700$ to
$0.625$, our temporal method from $0.475$ to $0.450$ at top-1 and from $0.775$ to $0.725$
at top-3, the LLM ranker from $0.625$ to $0.600$, the transfer test's Kendall $\tau$ from
$1.0$ to $0.667$, and the seed-to-seed spread to roughly a quarter of what it was. Two
findings survived and one appeared: the one-line rule still wins, mis-grouping stopped
looking worthless, and \emph{thp\_causal}'s advantage over its own control grew from two
incidents to seven.

What is still wrong is burst size, at $82$ alarms against $30$, and we have no mechanism
for it. The explanation we expected --- that a larger network gives one incident more
elements to affect --- is measurably false across the three backbones of
Section~\ref{sec:sweep}.

\paragraph{We shipped the answer in an alarm field.} Until 2026-10-21 every alarm carried
its role in the fault as \texttt{specificProblem}, so \texttt{specific\_problem = 'root'}
named a true root resource in all $28$ of the $40$ eval incidents where such an alarm
existed --- $1.000$ precision, $0.700$ top-1, from $59$ of $1{,}617$ alarms and no
reasoning. No number in this paper came from it: nothing in the benchmark or the baselines
read the field, and regenerating the dataset without it changed no scored rate, because the
fix draws no extra randomness and the two datasets differ in that one field. The cost was to
a reader, who could have taken the benchmark's best score with a single \texttt{WHERE}
clause. We report it rather than quietly reissuing the data, and the check that found it is
now part of the release (Section~\ref{sec:benchmark}). What it does not fix is the row below
it in Table~\ref{tab:leak}: a severity filter still scores $0.625$ --- the same as the best
method in Table~\ref{tab:results} --- and severity is a field the data is supposed to
have.

\paragraph{One realism finding here was previously wrong in our favour's opposite
direction.} Two of the three divergences this section reported before 2026-10-21 were
artefacts of measuring over the wrong alarm field, and the error made the generator look
worse than it is. The correction is in Section~\ref{sec:realism}; we record it here
because a resource paper's realism section is the part a reader cannot re-derive without
the data, and the field it is computed over is part of the measurement.

\paragraph{The excitation comparison is a proxy, not a fit.} What
Table~\ref{tab:realism} reports is a lift over base rate within a 300\,s window. No Hawkes
kernel is estimated, on either side, and the shape of the real excitation decay is
therefore unknown to us. A fitted model could change the sign of nothing in that table and
would still be the right measurement.

\section{Release}
\label{sec:release}
\paragraph{What ships, and where.} The generator is public, Apache-2.0, at
\url{https://github.com/samyama-ai/telecom-synthetic-data-generator}. It is a squashed
export carrying no development history, regenerated per release rather than maintained by
hand, so what is published is the state of a release and nothing about how it was reached.
The benchmark and the scoring scripts follow in a second repository; they are cited here as
forthcoming rather than listed as shipped, because a URL that does not resolve is worse
than none. What ships today is the generator, plus one
generated dataset per configuration. The dataset is not the artifact; the generator is.
A directory is a function of (pack version, config, seed) and the same seed reproduces its
files byte for byte, which is a tested property and not an aspiration --- so a reader who
disagrees with a number here can regenerate the input to it rather than download ours.
Every emitted directory carries its own \texttt{manifest.json} (seed, config, every input
and output with sha256) and \texttt{binding.json} (the vocabulary terms used and the build
files they came from).

\paragraph{Croissant.} Each directory carries MLCommons Croissant 1.0 metadata describing
its files, their checksums and the fields inside them, generated from that manifest so the
checksums in the metadata are the checksums of the files beside it rather than numbers
typed a second time. It names the ground-truth fields explicitly, including where each
alarm's mechanism lives, because a downloader who does not know which fields are the
answers will train on them.

\paragraph{Licence, settled.} The data is CC~BY~4.0, which carries the Internet Topology
Zoo's own terms for the backbone it is built on; the generator and the scoring code are
Apache-2.0. RFC~8632 and RFC~8345 are Simplified BSD and TMF642 is Apache-2.0. NORIA-O's
advertising clause, the one part that was open, is satisfied by the acknowledgement in
Section~\ref{sec:binding}: what we redistribute from NORIA-O is term names, not the
ontology, so the clauses binding a redistribution to retain its notice apply to the
generator's own vocabulary fixture --- where that notice now sits --- and not to the dataset.
The Croissant metadata asserts those two licences and carries the acknowledgement in a
generated field, so a release cannot drop it by editing prose. Until the review closed the
same metadata asserted no licence at all, which was the right default: the field a
downstream tool reads with no human present is the worst place to put a guess.

\paragraph{A DOI, and what this version does not yet carry.} The dataset is deposited as a
prepared, unpublished record: the archive, its checksum and the full deposit metadata exist,
and publishing it mints the DOI. That step is deliberately not taken in time for this
version, because a DOI is permanent and the record it names should be the one a venue will
review. Until then the resource is cited by this paper and by the public generator
repository above, which is what a reader needs to reproduce it; a revision will carry the
DOI. We state this rather than cite an
identifier that does not yet resolve.

\bibliographystyle{plainnat}
\bibliography{refs}

\begin{thebibliography}{15}
\providecommand{\natexlab}[1]{#1}
\providecommand{\url}[1]{\texttt{#1}}
\expandafter\ifx\csname urlstyle\endcsname\relax
  \providecommand{\doi}[1]{doi: #1}\else
  \providecommand{\doi}{doi: \begingroup \urlstyle{rm}\Url}\fi

\bibitem[rfc(2018)]{rfc8345}
A {YANG} data model for network topologies.
\newblock \url{https://www.rfc-editor.org/rfc/rfc8345}, 2018.

\bibitem[lem(2024)]{lemmarca2024}
{LEMMA-RCA}: A large multi-modal multi-domain dataset for root cause analysis.
\newblock arXiv preprint arXiv:2406.05375, 2024.
\newblock Microservice and water-treatment domains. CC BY-ND 4.0 on the site
  and in the paper; the Hugging Face header tags it cc-by-nc-4.0.

\bibitem[rca(2024)]{rcaeval2024}
{RCAEval}: A benchmark for root cause analysis of microservice systems.
\newblock arXiv preprint arXiv:2412.17015, 2024.
\newblock 735 failure cases over three microservice systems. Zenodo copy CC BY
  4.0, Hugging Face copy tagged MIT.

\bibitem[ope(2025)]{openrca2025}
{OpenRCA}: Can large language models locate the root cause of software
  failures?
\newblock ICLR 2025, 2025.
\newblock 335 cases over Bank, Market and Telecom domains. Cloud observability
  telemetry, not NOC alarms. MIT on the code; no licence stated on the data.

\bibitem[tna(2025)]{tnautorca2025}
{TN-AutoRCA}: Benchmark construction and agentic framework for telecom network
  root cause analysis.
\newblock arXiv preprint arXiv:2507.18190, 2025.
\newblock Announces TN-RCA530, 530 scenarios as knowledge graphs with
  per-incident labels. No repository, download, availability statement or
  dataset licence appears in the paper.

\bibitem[tel(2026)]{telerca2026}
{EvoCause}: Evolutionary causal discovery for root cause analysis in telecom
  networks.
\newblock arXiv preprint arXiv:2607.27290, 2026.
\newblock Introduces TeleRCA: 10,922 production incidents, 485,681 alarms,
  5,621 resources, 194 alarm types, expert \texttt{causeNode} annotation.
  States twice that TeleRCA has no ground-truth propagation graph. No URL in
  the paper.

\bibitem[{Huawei Noah's Ark Lab}(2026)]{gcastle_dataset}
{Huawei Noah's Ark Lab}.
\newblock gcastle-hub/dataset: alarm datasets from the {PCIC} 2021 causal
  discovery competition.
\newblock \url{https://github.com/gcastle-hub/dataset}, 2026.
\newblock MIT. The URL printed in \citet{qiao2023structural} no longer serves
  data; this mirror does. Subset \texttt{18V\_55N\_Wireless}: 34,838 alarms, 18
  types, 55 devices, 240 days of alarms, plus a type-level ground-truth causal
  graph and no per-incident labels. Subset \texttt{espk}: 141,853 alarms, a
  35-device topology and 49 incidents labelled (alarmed device set ->
  root-cause device), without timestamps. Accessed 2026-10-21.

\bibitem[{Huawei Paris Research Center, NetOp Team}(2025)]{telelogs}
{Huawei Paris Research Center, NetOp Team}.
\newblock {TeleLogs}.
\newblock \url{https://huggingface.co/datasets/netop/TeleLogs}, 2025.
\newblock Gated. Licence tag MIT while the card asks that the data not be
  shared or redistributed. 5G KPI/drive-test records, not alarms.

\bibitem[Knight et~al.(2011)Knight, Nguyen, Falkner, Bowden, and
  Roughan]{topologyzoo}
Simon Knight, Hung~X. Nguyen, Nickolas Falkner, Rhys Bowden, and Matthew
  Roughan.
\newblock The {Internet} {Topology} {Zoo}.
\newblock \emph{IEEE Journal on Selected Areas in Communications}, 29\penalty0
  (9):\penalty0 1765--1775, 2011.
\newblock CC BY 4.0.

\bibitem[Qiao et~al.(2023)]{qiao2023structural}
Jie Qiao et~al.
\newblock Structural hawkes processes for learning causal structure from
  discrete-time event sequences.
\newblock \emph{arXiv preprint arXiv:2305.05986}, 2023.
\newblock IJCAI 2023. Uses the Huawei PCIC 2021 alarm data; ground truth is an
  18x18 adjacency between alarm \emph{types}.

\bibitem[Tailhardat et~al.(2024)Tailhardat, Troncy, and Chabot]{noria2024}
Lionel Tailhardat, Rapha{\"e}l Troncy, and Yoan Chabot.
\newblock {NORIA-O}: An ontology for anomaly detection and incident management
  in {ICT} systems.
\newblock In \emph{ESWC}, 2024.
\newblock BSD-4-Clause with the advertising clause. Alarm vocabulary aligns to
  ITU-T X.733/X.721 and TMF638; it references neither RFC 8632 nor TMF642.

\bibitem[Tailhardat et~al.(2026)Tailhardat, Troncy, Chabot, Ramparany, Folz,
  and Kavanagh]{noria_draft}
Lionel Tailhardat, Rapha{\"e}l Troncy, Yoan Chabot, Fano Ramparany, Pauline
  Folz, and Anne Kavanagh.
\newblock Knowledge graphs for enhanced cross-operator incident management and
  network design.
\newblock Internet-Draft draft-tailhardat-nmop-incident-management-noria-05,
  2026.
\newblock Updated 2026-08-07, expires 2027-02-08. Cites RFC 8345; no occurrence
  of RFC 8632 or TMF642.

\bibitem[{TM Forum}(2025)]{tmf642}
{TM Forum}.
\newblock {TMF642} alarm management {API}, 2025.
\newblock v5.0, directory release 2025-01-23. Apache-2.0.

\bibitem[Vallin and Bjorklund(2019)]{rfc8632}
Stefan Vallin and Martin Bjorklund.
\newblock A {YANG} data model for alarm management.
\newblock \url{https://www.rfc-editor.org/rfc/rfc8632}, 2019.

\bibitem[Zhang et~al.(2021)Zhang, Kalander, et~al.]{zhang2021influence}
Keli Zhang, Marcus Kalander, et~al.
\newblock An influence-based approach for root cause alarm discovery in telecom
  networks.
\newblock \emph{arXiv preprint arXiv:2105.03092}, 2021.
\newblock States data and code are available at github.com/shaido987/alarm-rca;
  that repository contains a single 378-byte README and no licence.

\end{thebibliography}

\end{document}